\documentclass[8.5pt,twoside,twocolumn]{article}
\usepackage[super,sort&compress,comma]{natbib} 
\usepackage[version=3]{mhchem}
\usepackage{bm}
\usepackage{times}
\usepackage{sectsty}
\usepackage{balance} 

\usepackage{graphicx}
\usepackage{lastpage}
\usepackage[format=plain,justification=raggedright,singlelinecheck=false,font=small,labelfont=bf,labelsep=space]{caption} 
\usepackage{fancyhdr}
\begin{document}

\thispagestyle{plain}
\fancypagestyle{plain}{
\renewcommand{\headrulewidth}{1pt}}
\renewcommand{\thefootnote}{\fnsymbol{footnote}}
\renewcommand\footnoterule{\vspace*{1pt}%
\hrule width 3.4in height 0.4pt \vspace*{5pt}} 
\setcounter{secnumdepth}{5}

\makeatletter 
\def\subsubsection{\@startsection{subsubsection}{3}{10pt}{-1.25ex plus -1ex minus -.1ex}{0ex plus 0ex}{\normalsize\bf}} 
\def\paragraph{\@startsection{paragraph}{4}{10pt}{-1.25ex plus -1ex minus -.1ex}{0ex plus 0ex}{\normalsize\textit}} 
\renewcommand\@biblabel[1]{#1}            
\renewcommand\@makefntext[1]%
{\noindent\makebox[0pt][r]{\@thefnmark\,}#1}
\makeatother 
\renewcommand{\figurename}{\small{Fig.}~}
\sectionfont{\large}
\subsectionfont{\normalsize} 

\fancyfoot{}
\fancyfoot[RO]{\footnotesize{\sffamily{1--\pageref{LastPage} ~\textbar  \hspace{2pt}\thepage}}}
\fancyfoot[LE]{\footnotesize{\sffamily{\thepage~\textbar\hspace{3.45cm} 1--\pageref{LastPage}}}}
\fancyhead{}
\renewcommand{\headrulewidth}{1pt} 
\renewcommand{\footrulewidth}{1pt}
\setlength{\arrayrulewidth}{1pt}
\setlength{\columnsep}{6.5mm}
\setlength\bibsep{1pt}

\twocolumn[
  \begin{@twocolumnfalse}
\noindent\LARGE{\textbf{Interactions and design rules for assembly of porous colloidal mesophases$^\dag$}}
\vspace{0.6cm}

\noindent\large{\textbf{Beth A. Lindquist,\textit{$^{a}$} Sayantan Dutta,\textit{$^{a\ddag}$} Ryan B. Jadrich,\textit{$^{a}$} Delia J. Milliron,\textit{$^{a}$} and Thomas M. Truskett$^{\ast}$\textit{$^{a}$}}}\vspace{0.5cm}

\noindent \normalsize{Porous mesophases, where well-defined particle-depleted `void' spaces are present within a particle-rich background fluid, can be self-assembled from colloidal particles interacting via isotropic pair interactions with competing attractions and repulsions. While such structures could be of wide interest for technological applications (e.g., filtration, catalysis, absorption, etc.), relatively few studies have investigated the interactions that lead to these morphologies and how they compare to those that produce other micro-phase-separated structures, such as clusters. In this work, we use inverse methods of statistical mechanics to design model isotropic pair potentials that form porous mesophases. We characterize the resulting porous structures, correlating features of the pair potential with the targeted pore size and the particle packing fraction. The former is primarily encoded by the amplitude and range of the repulsive barrier of the designed pair potential and the latter by the attractive well depth. We observe a trade-off with respect to the packing fraction of the targeted morphology: greater values support more spherical and monodisperse pores that themselves organize into periodic structures, while lower values yield more mobile pores that do not assemble into ordered structures but remain stable over a larger range of packing fraction. We conclude by commenting on the limitations of targeting a specific pore diameter within the present inverse design approach as well as by describing future directions to overcome these limitations.}
\vspace{0.5cm}
 \end{@twocolumnfalse}
  ]

\footnotetext{\textit{$^{a}$~McKetta Department of Chemical Engineering, University of Texas at Austin, Austin, Texas 78712, USA. E-mail: truskett@che.utexas.edu}}

\footnotetext{\ddag~Present address: \textit{Department of Chemical Engineering, Indian Institute of Technology, Kharagpur, West Bengal, India-721302}}

\section{Introduction}

Colloidal materials are well-suited to act as building blocks for self-assembled structures because their interparticle interactions can be readily tuned through colloid shape and composition, passivating ligand identity, salt concentration, solvent quality, etc.~\cite{CollInt1} Anisotropy is often used in order to direct the assembly process. For instance, exotic structures of particles with highly directional, anisotropic interactions--so-called patchy particles~\cite{patchy_review1,patchy_review2}--have been observed to form a kagome lattice in two dimensions,~\cite{kagome} as well as well-defined clusters that form a variety of three dimensional shapes.~\cite{DNA_clusters} Further, particles containing both patchy attractions as well as shape anisotropy have been found to self-assemble into capsid-like structures under certain conditions.~\cite{capsid} Finally, janus particles have been shown~\cite{janus}, in analogy to diblock copolymers~\cite{BCPs}, to self-assemble into a variety of micro-phase-separated states, such as tubes, columns and lamellar sheets. 

Interestingly, anisotropic interactions are \emph{not} required to assemble mesoscopic structures. For example, it is well established that isotropic pair interactions that exhibit competing attractions and repulsions can form micro-phase-separated states,~\cite{pores,SALR1,SALR3,mod_phases1,mod_phases2,mod_phases3,mod_phases4,mod_phases5} and a variety of recent studies have focused on characterizing phases formed at lower particle concentration, including clusters, columns, and lamellar sheets.~\cite{ryanPRE,jon1,jon2,clusters1,clusters2,2D_salr_stripes,DFT_cluster_lamellar,columns_lamellae,CSD_1,SALR2,wu_lowrho,clusters3} However, it has also been theoretically predicted that, at higher particle concentrations, there exist analogues to clusters and columns where the particles and the void space are inverted, yielding a background fluid that is excluded from columnar or spherical regions of space.~\cite{postulated_phases_sear,mean_field_assembly_0,mean_field_assembly_1,mean_field_assembly_2,2D_microphase,lattice_bubbles_1} This equivalence between particle-filled and particle-depleted space has been confirmed with simulation.~\cite{pores,simulated_phases,lattice_bubbles_2} However, the correspondence between pair interactions and resulting porous mesophases of `inverse clusters' is significantly less understood than that for particle clusters, particularly in three dimensions. 

Porous mesophases provide a self-assembled alternative to molecular cages, such as zeolites, and therefore could find application in size-selective solubility, filtration, or catalysis due to the presence of a percolated fluidic network possessing large exposed surface areas. However in order to realize such structures, the connection between interparticle interaction and the resulting porous structural properties must be more fully understood. We recently demonstrated that isotropic pair potentials that form structured fluids with pores of a prescribed size can be designed via inverse methods of statistical mechanics.~\cite{pores} Inverse design strategies have proven to be powerful tools for the discovery of pair interactions that self-assemble into various complex architectures;~\cite{ST_inv_des_review,AJ_inv_des_review,ID_crystals_TS,ID_crystals, ID_clusters,BL_RE} in this work, we again leverage inverse design to systematically optimize for a variety of pore-forming pair potentials at different particle concentrations and targeted pore sizes in both two and three dimensions. To gain insights into the design rules for these materials, we explore the relationships between the properties of porous mesophases and key features of the interparticle interactions that drive their assembly. 

This paper is organized as follows. In Section~\ref{sec:methods}, we discuss our approach for inverse design as well as other computational considerations. We present our simulation results in Section~\ref{sec:results}, showing that fluid mesophases displaying porous architectures can indeed be assembled for a variety of targeted pore sizes and particle concentrations and that these properties correlate with features of the designed pair potential. Finally, we conclude in Section~\ref{sec:conclusions}.

\section{Computational Methods}
\label{sec:methods}

\subsection{Iterative Boltzmann Inversion}
\label{subsec:IBI}

The Henderson uniqueness theorem states that if there exists a dimensionless pair potential $\beta u(r)$ [where $\beta\equiv 1/k_{\text{B}}T$, $k_{\text{B}}$ is Boltzmann's constant, and $T$ is temperature] with equilibrium configurations that produce a given radial distribution function $g(r)$, then it is unique in being able to do so.~\cite{HansenMcDonald,Henderson} Iterative Boltzmann Inversion (IBI) provides a simple, heuristic framework to find the $\beta u(r)$ which, at equilibrium, would produce a structural ensemble with a radial distribution function equal to that of a targeted ensemble of configurations, $g_{\text{tgt}}(r)$.~\cite{IBI1,IBI2,IBI3,test_of_ID_schemes} In keeping with our prior work,~\cite{pores} we generate $g_{\text{tgt}}(r)$ for porous mesophases from a simulation of a binary mixture of mobile hard-core-like particles with a diameter of $\sigma$ in the presence of a periodically replicated frozen template of 32 larger spheres placed on a face-centered cubic (FCC) lattice in three dimensions (3D) or a triangular lattice in two dimensions (2D). The template serves to exclude the smaller particles from spherical regions of space, thereby ensuring that their pair correlations are consistent with the desired porous structure. Accordingly, the diameter of the frozen spheres dictates the pore size. Previously,~\cite{pores} we chose a pore size of $4\sigma$ and targeted a single particle packing fraction. In this work, we also design mesophases with pore diameters of $3\sigma$ and $5\sigma$ at different packing fractions. We found in prior work~\cite{pores} that $\beta u(r)$ optimized from a target simulation with a lattice constant of $\approx10.5\sigma$, or a nearest neighbor distance of $7.4\sigma$, was able to successfully form a porous mesophase. For the different pore sizes considered here, we use the same ratio of nearest-neighbor distance to pore diameter (1.85) to fabricate the template in the target simulations.

With $g_{\text{tgt}}(r)$ in hand, we can carry out the IBI optimization. First, a single-component simulation is performed, where the particles interact via $u^{(i)}(r)$, where $i$ indicates the step of the optimization. From this simulation, $g^{(i)}(r)$ is calculated, and $u^{(i)}(r)$ is updated according to the following relation:
\begin{equation} \label{eqn:ibi_equation}
u^{(i+1)}(r)\equiv u^{(i)}(r)+\alpha k_{\text{B}}T \textnormal{ln}\bigg[\dfrac{g^{(i)}(r)}{g_{\text{tgt}}(r)}\bigg]
\end{equation}
where $\alpha$ is the step size for the potential update. The initial guess for the potential is given by direct Boltzmann inversion, i.e., $u^{(1)}(r)\equiv-k_{\text{B}}T\textnormal{ln}[g_{\text{tgt}}(r)]$.~\cite{IBI1} The update scheme is physically motivated in that if the simulated $g^{(i)}(r)$ is over-coordinated relative to $g_{\text{tgt}}(r)$ at a given interparticle distance, $r^{*}$, then $u(r^{*})$ is updated to be more repulsive, and vice versa.

A cut-off for the potential ($r_{\text{c}}$) is needed in the IBI framework; in prior work,~\cite{pores} we found that a single attractive well, coupled with a longer-ranged repulsive barrier (`hump'), can generate porous architectures. Therefore, our general strategy to determine the cut-off is to perform IBI with a longer cut-off than needed and find the minimum ($r_{\text{c}}$) in the potential following the repulsive hump. The potential is then cut at that minimum and shifted by a constant such that $u(r_{\text{c}}) = 0$, and it is subsequently re-optimized with the new cut-off until convergence. Preliminary calculations indicated that the optimal $r_{\text{c}}$ is approximately equal to the nearest-neighbor distance between the large spheres of the template in the target simulation, though in 2D we sometimes found that a slightly larger cut-off was needed. 

The target simulation used a steeply repulsive Weeks-Chandler-Andersen (WCA) potential (derived from a 50-25 generalized Lennard-Jones model) as an approximation to the hard-sphere interaction.~\cite{HansenMcDonald} The target simulations and those required for the iterative procedure were carried out in either Gromacs versions 4.6.5 or 5.0.6~\cite{GROMACS_1,GROMACS_2} in reduced units such that $\beta \epsilon_{WCA}=1$, the mass, $m=1$, and the diameter $\sigma=1$ with a time step of $dt=0.001\sqrt{\sigma m/k_{\text{B}}T}$. Simulations were performed in the canonical ensemble using periodic boundary conditions and a velocity-rescaling thermostat with a time constant of $100dt$. IBI calculations were carried out with the Versatile Object-oriented Toolkit for Coarse-graining Applications (VOTCA) package,~\cite{VOTCA_1,VOTCA_2} which interfaces with Gromacs. Simulations were visualized using Visual Molecular Dynamics (VMD).~\cite{VMD}

\subsection{Characterization}
\label{subsec:char}

In order to analyze the porous architectures, we randomly inserted test spheres (circles in 2D) of diameter $2\sigma$ into configurations from the simulations, keeping those that do not overlap with the cores of the fluid particles (taken to be of diameter $\sigma$). Discrete void regions were identified by performing a clustering analysis of the inserted spheres in the typical way,~\cite{CSD_1,CSD_2} where overlapping test spheres are taken to be connected neighbors. The volumes (areas in 2D) were computed via Monte Carlo integration. The resultant volumes (areas) are plotted here in terms of the diameter of an equi-volume sphere (equi-area circle). 

In order to detect transitions between various microphases as a function of particle concentration, the relative anisotropy factor,~\cite{kappa2}
\begin{equation} \label{eqn:k2}
\kappa^{2}=\dfrac{3}{2} \dfrac{\lambda_{x}^{4}+\lambda_{y}^{4}+\lambda_{z}^{4}}{(\lambda_{x}^{2}+\lambda_{y}^{2}+\lambda_{z}^{2})^{2}} - \dfrac{1}{2}
\end{equation}
where $\lambda^{2}_{x}$, $\lambda^{2}_{y}$, and $\lambda^{2}_{z}$ are the eigenvalues of the gyration tensor, was calculated for both the pores--as defined by the inserted test spheres described above--and for the particles themselves. A $\kappa^{2}$ value of 0 indicates shapes that are symmetrically distributed over the three coordinate axes, e.g., as for a sphere; by contrast, a line has a relative anisotropy factor of 1. For the clustering analysis of particles, the cut-off distance for neighbor determination was defined as $r_{\text{csd}}=1.15\sigma$. The gyration tensor of the clusters is calculated from the particle centers as is typical for particle-based systems. For voids we do not directly use the centers of the inserted test particles, as we wish to treat the void as a continuous volume. Thus, we recast each discrete region of void space as a grid with a spacing of $0.5\sigma$, i.e., the grid points are accepted if they overlap with a given aggregate of inserted test particles. Performing the calculation in this way approximates the volume-integrated calculation of $\kappa^{2}$ for voids. For both particles and void space, these calculations are complicated by the co-existence of the primary structural features with ``monomer'', literal monomers or other small clusters for the former and individual test spheres for the latter. Small clusters have very large variance in $\kappa^{2}$ due to the particle size and cluster size being similar in scale, and void spaces that are very near to the inserted test sphere size are necessarily nearly spherical and therefore trivially have very small relative anisotropy factors. To determine $\kappa^{2}$ for the primary structural features while minimizing interference from very small clusters or pores, we only include clusters that are larger than 8 particles in size (any preferred size scales apparent in the cluster size distributions were always greater than 8 particles in size), and we also omitted any pores with an effective diameter (diameter of an equi-volume sphere), $d_{\text{eff}}$, less than 2.1$\sigma$. 
Average $\kappa^{2}$ values were weighted by either the number of particles in the cluster or volume of the void regions,
yielding a metric that reports on the anisotropy of the structures that constitute the bulk of the particle-rich (or particle-poor) space. This weighting also lends some flexibility to the exact values of the cut-offs listed above, helpful since the optimal values vary somewhat as a function of $\eta$. 

\section{Results and Discussion}
\label{sec:results}

In previous work,\cite{pores} we optimized a pair potential to form an ordered porous 3D mesophase with pore diameter $d_{\text{pore}}=4\sigma$ and particle packing fraction $\eta_{\text{opt}} \equiv \pi N \sigma^{3}/6V=0.31$, where $N$ is the number particles and $V$ is the volume. Here, we extend this work to include $d_{\text{pore}} = 3\sigma$, $4\sigma$, and $5\sigma$ and $\eta_{\text{opt}}=0.22$, 0.26, 0.31, and 0.35. We also optimized for pores of the same diameters in 2D at $\eta_{\text{opt}} \equiv \pi N \sigma^{2}/4A = 0.60$ ($A$ is the area). In all cases, matching between the radial distribution functions computed from the target simulation and the optimized potential was excellent, as can be seen in Fig. A1 of the Appendix. From these optimizations, we explore the relationship between these parameters and key features of the pair interactions.

\subsection{Effect of Varying Pore Diameter}
\label{subsec:pore_size}

\begin{figure}[!htb]
  \includegraphics{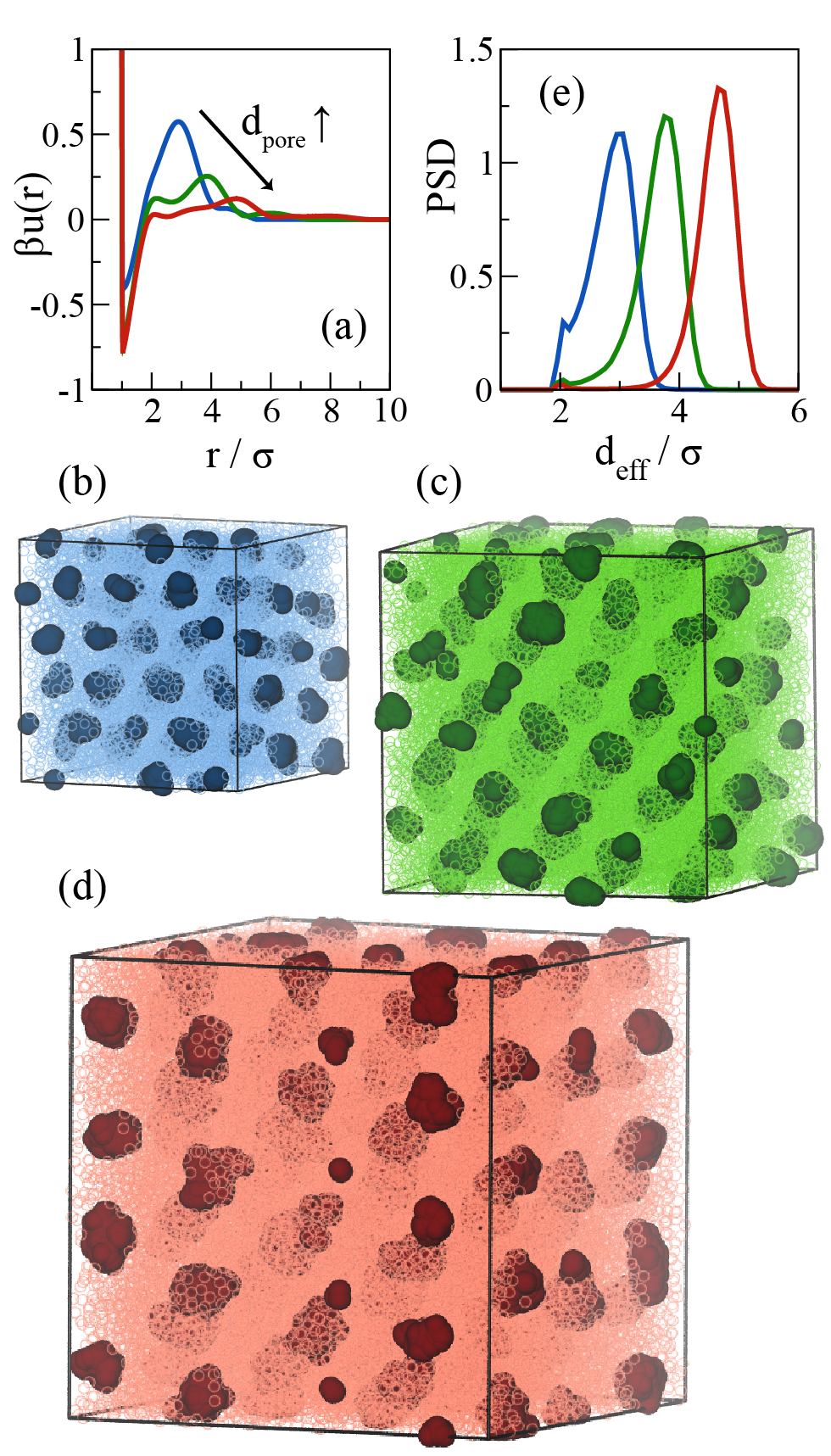}
  \caption{For $\eta_{\text{opt}}=0.31$, (a) the dimensionless optimized pair potentials, $\beta u(r)$, (b-d) representative configurations of particles (transparent, lighter) and inserted test spheres (opaque, darker), and (e) pore size distribution (PSD) functions for $d_{\text{pore}}=3$ (blue), 4 (green), and 5 (red).}
  \label{fgr:ps}  
\end{figure}

The pair potentials optimized for systems with $\eta_{\text{opt}}=0.31$ and $d_{\text{pore}}=3\sigma$, $4\sigma$, or $5\sigma$ are shown in Fig.~\ref{fgr:ps}a. All can be qualitatively described as possessing an attractive well and a longer-ranged repulsive hump. The maximum in the repulsive hump falls close to the targeted pore diameter. In prior work,\cite{pores} we rationalized this finding by noting that the optimized potentials were most faithful to the target simulation when the pore diameter was commensurate with the extent of the particle-occupied space separating the pores. When the pore diameter is directly encoded in the pair potential (as in Fig.~\ref{fgr:ps}a) and the pore diameter and pore-pore spacing are matched in this fashion, the primary repulsive feature of the interaction--emanating from the particle center isotropically as a spherical shell--is positioned such that it, while necessarily raising the energy of other particles in the surrounding `matrix', helps to stabilize (i.e., evacuate particles from) multiple pores. This slaving of the optimal thickness of the particle-rich regions to the pore diameter motivated our choice of 1.85 for the pore nearest-neighbor distance to pore diameter ratio (discussed in Sect.~\ref{subsec:IBI}), as it allows for the thickness of the particle-rich layer to be about equal in length to the pore diameter for much of the simulation box.  

As the pore size of the targeted mesophase grows, the increased range of the repulsive hump results in more particles at the perimeter contributing to the stabilization of the void space; therefore, each individual particle contributes to a lesser degree. As a result, the optimal potential can afford a noticeable reduction in the amplitude of this repulsive feature with increasing $d_{\text{pore}}$. On the whole, for this potential form, the length scales associated with the size of the microphase structures are directly encoded into the pair interaction. Interestingly though, the attractive well is much less sensitive than the repulsive hump to $d_{\text{pore}}$, though there is some deviation for $d_{\text{pore}}=3$, discussed below.

In addition to this coarse-grained description, there are other smaller scale features in the potential. We interpret these features as a result of ``over-fitting'' the optimized $g(r)$ to that of the target. Since the target simulation included only excluded volume effects, the local structuring is not representative of a model with the type of shorter-ranged attractions that the optimized potentials possess. Thus, humps at multiples of the particle size, particularly at 2$\sigma$, may function to ``space out'' particle aggregates to more closely mimic the local environment present in the target configurations. This feature at $2\sigma$ is somewhat muted for $d_{\text{pore}}=3$, as it begins to merge with the repulsive hump. The encroachment of the repulsive hump into smaller interparticle distances may necessitate the somewhat more shallow attractive well in this case to keep the particles apart, as observed in Fig.~\ref{fgr:ps}a. In general though, the distinction between sticky and hard core-like particles is not essential to mesoscopic pore formation, which is corroborated by our prior work~\cite{pores} showing that the smaller scale minima, maxima, and shoulders (which develop in addition to the primary attractive and repulsive component) are not essential to form pores. 

\begin{figure}[!htb]
  \includegraphics{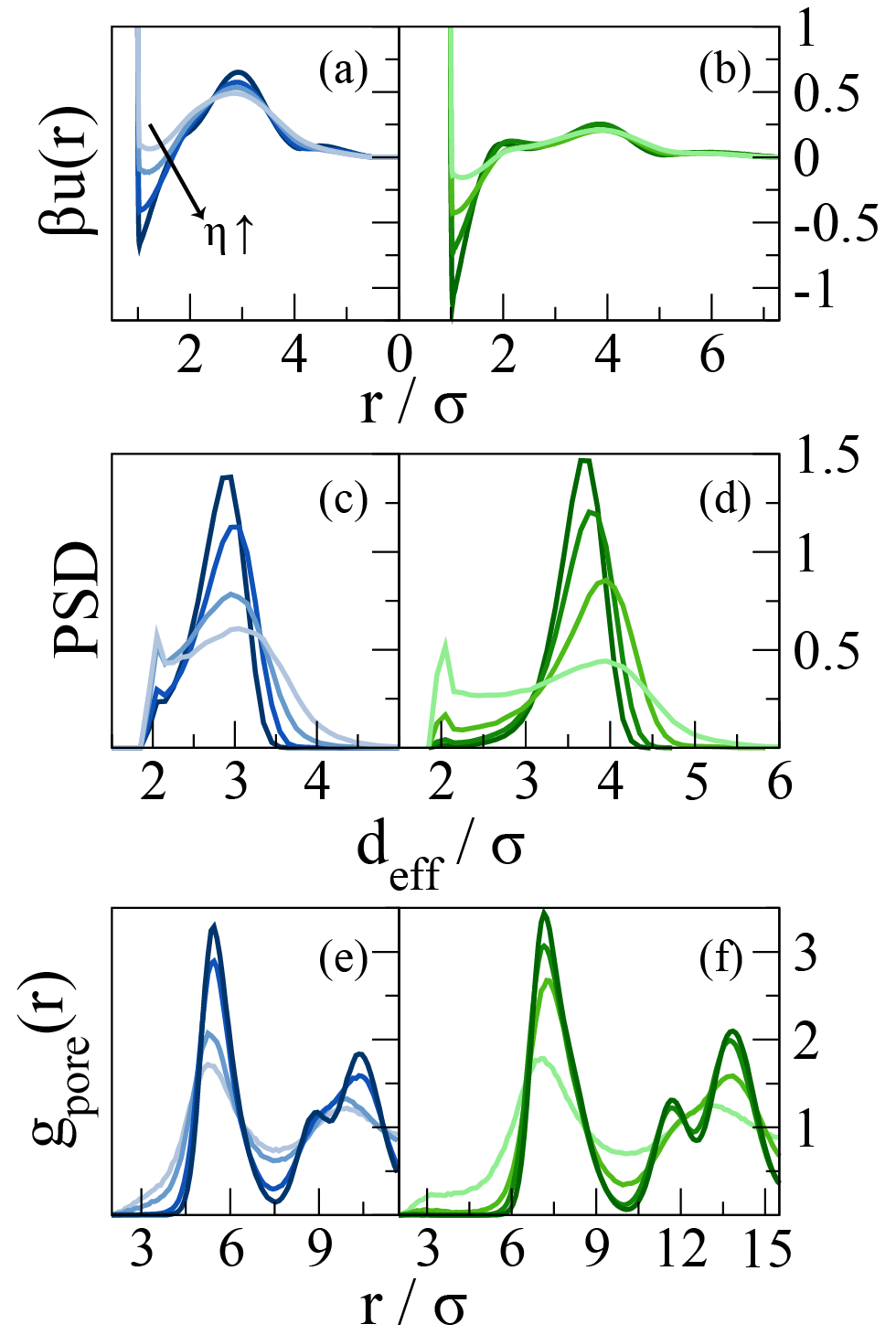}
  \caption{For $d_{\text{pore}}=3$ (blue, left) and 4 (green, right), (a,b) optimized dimensionless pair potentials, $\beta u(r)$, (c,d) pore size distribution (PSD) functions, and (e,f) $g(r)$ of the pore centers for $\eta_{\text{opt}}=0.22$, 0.26, 0.31, and 0.35 (from light to dark).}
  \label{fgr:density}   
\end{figure}

In Fig.~\ref{fgr:ps}b-d, we show representative configurations resulting from simulation of $\beta u(r)$, including the inserted test spheres that do not overlap with the particle cores. The simulated particles are shown as lighter, more transparent spheres, and the inserted spheres in the voids are represented as darker, opaque volumes. The presence of ordered and relatively spherical void regions, i.e., pores, is readily apparent upon visual inspection for all $d_{\text{pore}}$. For a quantitative analysis, the pore size distributions (PSDs) are shown in Fig.~\ref{fgr:ps}e, where it can be seen that all potentials faithfully reproduce the targeted pore size--the maxima are within 0.35$\sigma$ of the designed size. There is clearly some mild polydispersity present; the standard deviation of the PSDs are 0.36$\sigma$, 0.34$\sigma$, and 0.29$\sigma$ (ordered by increasing $d_{\text{pore}}$) upon fitting the PSD to a Gaussian function. As described in Section~\ref{subsec:char}, the pores are identified by performing a clustering analysis on the inserted test spheres. The small feature on the left side of the distribution is an artifact of the finite diameter of the inserted test spheres (2$\sigma$). In the Appendix, we show that $\approx 2\sigma$ is the optimal diameter for the test sphere and explore the dependence of the PSD on the diameter of the test spheres.  

\subsection{Effect of Varying Packing Fraction}

In the previous section, we noted that, while the repulsive hump seems to explicitly encode $d_{\text{pore}}$, the attractive well has a less direct dependence on the pore size. However, Fig.~\ref{fgr:density}a,b show that the attractive well responds sensitively and in a systematic fashion to modulation of $\eta_{\text{opt}}$, becoming progressively shallower at lower $\eta_{\text{opt}}$. In the target simulation, the pores occupy equal volumes of space, independent of $\eta_{\text{opt}}$. So as $\eta_{\text{opt}}$ increases, the particles must be closer together on average, necessitating stronger inter-particle attractions. In order to fortify the pore boundaries against this increased particle stickiness, the magnitudes of the repulsions also grow weakly. 

In Fig.~\ref{fgr:density}c,d, we see that the PSDs are all centered roughly around the prescribed diameter, but the variance in the distribution increases for the lower $\eta_{\text{opt}}$ values. Fig.~\ref{fgr:density} shows the results for $d_{\text{pore}}=3 \sigma$ and 4$\sigma$, with the analogous results for $d_{\text{pore}}=5\sigma$ given in the Appendix. Presumably as a result of both the increased density and decreased polydispersity at the higher target $\eta$ values, the pores at lower $\eta_{\text{opt}}$ are organized in a disordered fashion and the pores at higher $\eta_{\text{opt}}$ are crystalline on a body-centered cubic (BCC) lattice. Note that, as discussed previously,\cite{pores} the BCC pore lattice is likely stabilized over the targeted (and closely related) FCC lattice morphology in the designed systems because pores in the assembled structures have more diffuse interfaces than those of the template, yielding softer effective pore-pore interactions that favor the BCC morphology. This softness in interactions as well as the onset of the split peak in the second coordination shell in the $g(r)$ between pore centers (characteristic of crystallization) at the highest $\eta_{\text{opt}}$ can be seen in Fig.~\ref{fgr:density}e,f, where only voids comprising three or more inserted test spheres are included.   

\begin{figure}[!htb]
  \includegraphics{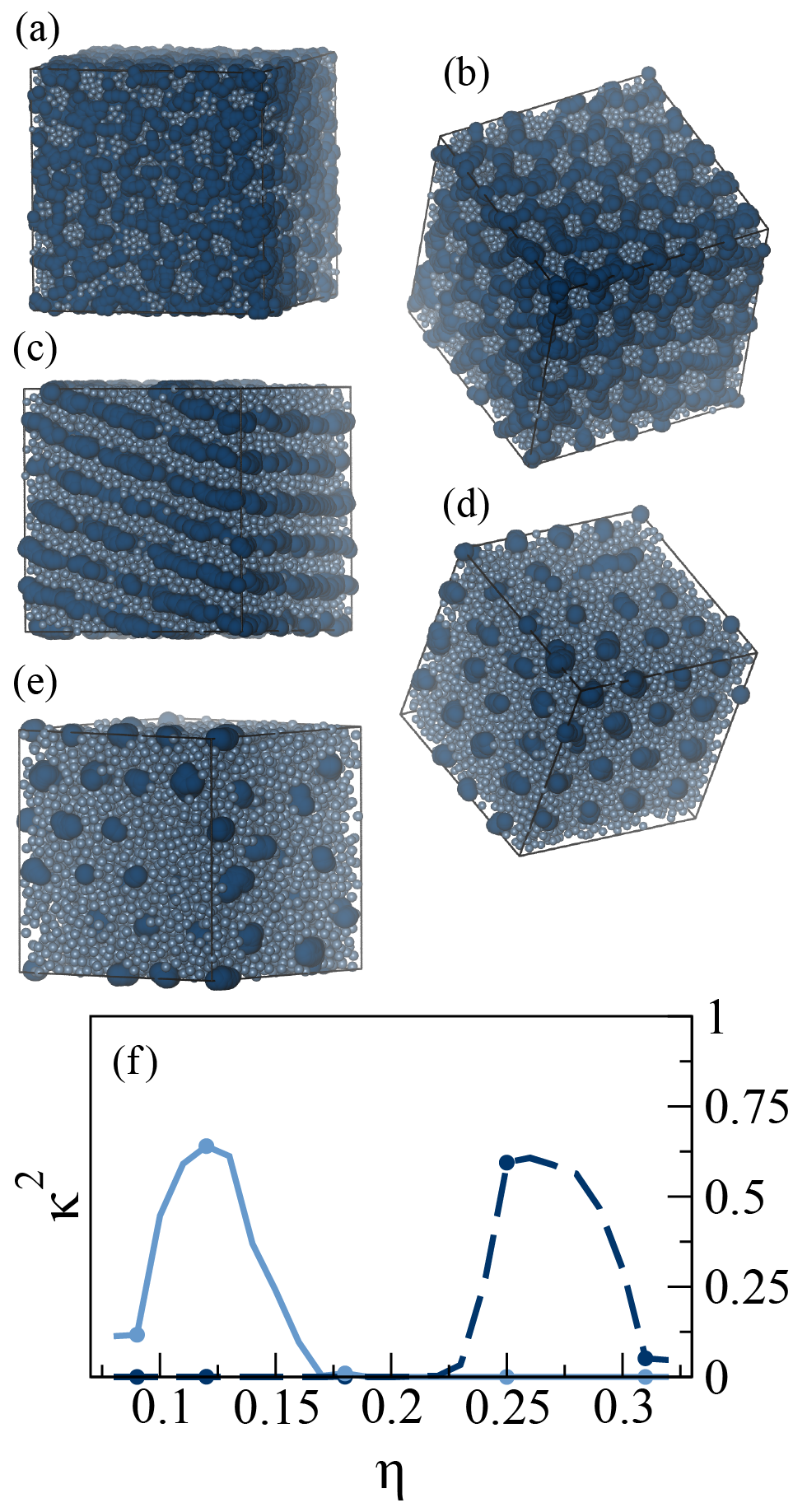}
  \caption{For potentials optimized for $d_{\text{pore}}=3$ and $\eta_{\text{opt}}=0.31$, (a-e) configurations at various values of $\eta$, corresponding to clusters, columns of particles, lamellar sheets, columns of voids, and pores, respectively, and (f) the relative anisotropy factor $\kappa^{2}$ as a function of $\eta$ for the particles (solid lines) and the volumes defined by the inserted test spheres (dashed lines). The dots on the curves correspond to the volume fractions of panels (a-e), from left to right. In the visualizations, the simulated particles are lighter in color and the inserted test spheres are darker.}
  \label{fgr:k35}   
\end{figure}

\begin{figure}[!htb]
  \includegraphics{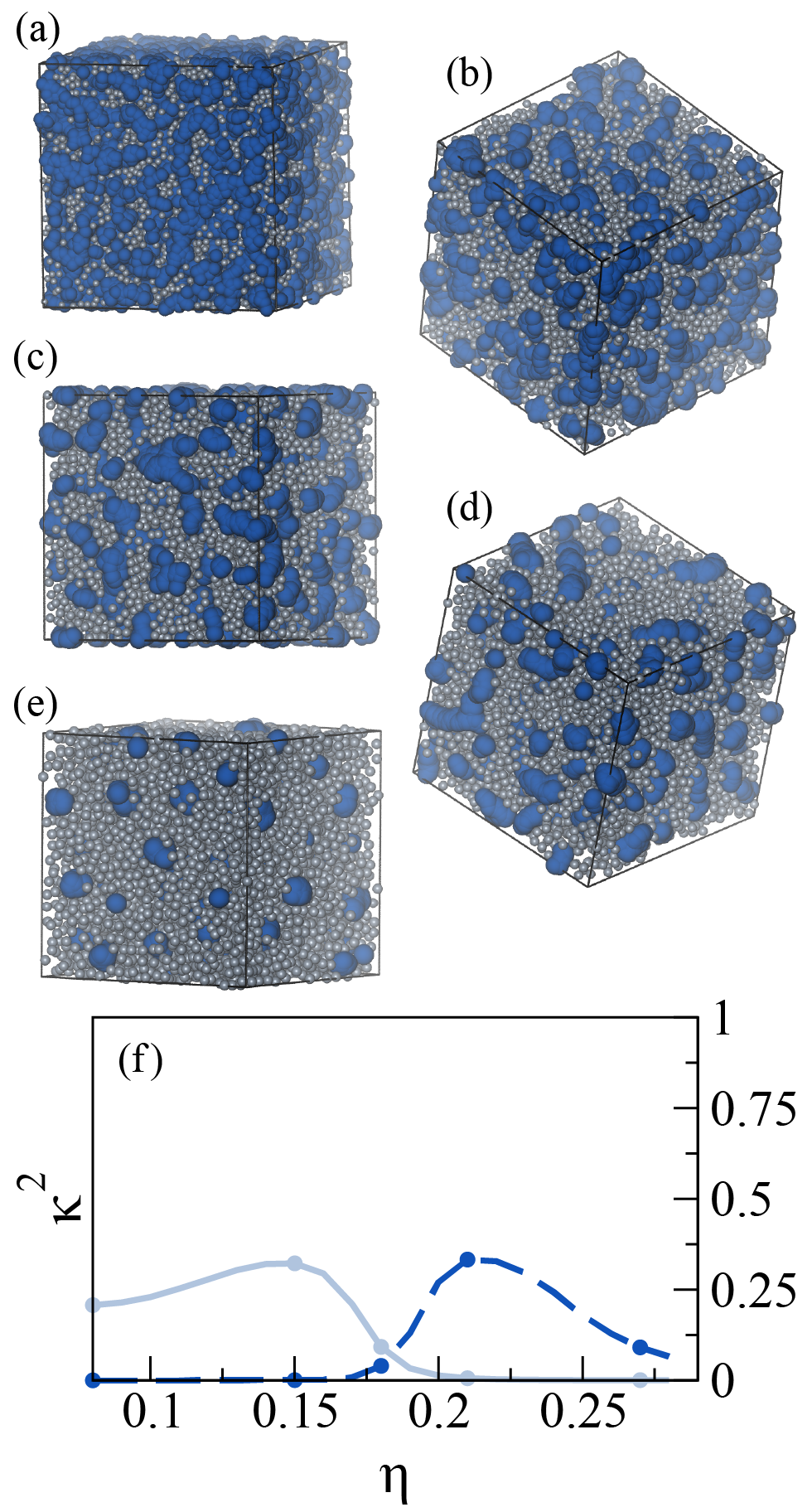}
  \caption{For potentials optimized for $d_{\text{pore}}=3$ and $\eta_{\text{opt}}=0.26$, (a-e) configurations at various values of $\eta$, beginning with clusters in panel (a) and ending at pores in panel (e), with a range of relatively amorphous, unstructured configurations in between these phases, and (f) the relative anisotropy factor $\kappa^{2}$ as a function of $\eta$ for the particles (solid lines) and the volumes defined by the inserted test spheres (dashed lines). The dots on the curves correspond to the volume fractions of panels (a-e), from left to right. In the visualizations, the simulated particles are lighter in color and the inserted test spheres are darker.}
  \label{fgr:k30}   
\end{figure}

To test the sensitivity of the assembled pore morphology on particle packing fraction for different design conditions, we carry out a series of simulations with pair potentials designed at various $\eta_{\text{opt}}$ for a range of $\eta$.  
We observe two qualitatively distinct progressions in phase behavior upon decrease in $\eta$ depending on whether the porous mesophase was designed with lower or higher $\eta_{\text{opt}}$. For systems optimized at higher packing fractions ($\eta_{\text{opt}}=0.31$ and 0.35) and $d_{\text{pore}}=3$, we find that decreasing $\eta$ causes pores to first merge forming inverse columns (i.e., columns of void space) followed by lamellar sheets, then columns of particles, and finally clusters of particles. Representative snapshots of the particles and the inserted overlapping spheres characterizing the pores are shown in Fig.~\ref{fgr:k35}a-e for $\eta_{\text{opt}}=0.31$. This behavior is entirely consistent with the results of the one system examined in our prior work ($\eta_{\text{opt}}=0.31$, $d_{\text{pore}}=4$).~\cite{pores}  In Fig.~\ref{fgr:k35}f, we show the relative anisotropy factor, $\kappa^{2}$, of the microphase structures as a function of $\eta$. At and near $\eta_{\text{opt}}$, approximately spherical pores exist, and the particles are percolated and relatively homogeneous (though absent from the void regions). Accordingly, $\kappa^{2}$ for the pores is rather small but non-zero, whereas the $\kappa^{2}$ for the particles is effectively zero. The latter finding indicates that the particles are not only percolated but also spread over all three dimensions of the box. As the packing fraction is decreased, the pores become more anisotropic eventually fully extending into columns (see Fig.~\ref{fgr:k35}d). In the clustering analysis, sometimes multiple columns come into contact with one another and are treated as a single column. This leads to a lower computed relative anisotropy factor than for a single column, which has the effect of lowering $\kappa^{2}$ on average. Correlations of cluster/pore size and $\kappa^{2}$ are shown in the Appendix to illustrate this point. 

Upon additional decrease in $\eta$, the inverse columns transition to a lamellar regime;
see, for example, the configuration in Fig.~\ref{fgr:k35}c. This corresponds to a sharp decrease in $\kappa^{2}$ to very small, but non-zero, values for the void space. $\kappa^{2}$ values for the particles are similar in magnitude. Individual lamellar sheets would not be expected to possess such a low relative anisotropy factor. However, the facile exchange of particles between sheets (evident from visualization of the trajectory data), coupled with the large interfacial areas yet relatively small distances between features, renders a pairwise, distance-based clustering analysis unable to identify single lamellar sheets. As a result, many (sometimes even all) of the sheets are clustered into a single object, which has a much lower $\kappa^{2}$ value than a single sheet. Nonetheless, visual inspection confirms that a computed $\kappa^{2}$ value of near zero for both particles and voids indeed corresponds to a lamellar regime; therefore, the $\kappa^{2}$ calculations provide an accurate location for the mesophase boundaries, despite the limitations of the analysis in this regime.

Upon further decreasing $\eta$, $\kappa^{2}$ associated with the particles grows sharply, indicating column formation (Fig.~\ref{fgr:k35}b); commensurately, the void space percolates in all dimensions, yielding $\kappa^{2}$ values associated with the test spheres (void region) of zero. Finally, from columns of particles, further reduction in $\eta$ yields comparatively spherical clusters as indicated by the drop in $\kappa^{2}$. These clusters are relatively small ($\approx 20$ particles in size), and so they are not perfectly spherical (Fig.~\ref{fgr:k35}a). If we compare the packing fractions at which these transitions occur to our prior work (performed at $d_{\text{pore}}=4$)\cite{pores}, grouping any lamellar and bi-continuous phases together, we see that the transitions occur within 0.01 of the same packing fractions, despite the different pore sizes. Therefore, it appears that $\eta_{\text{opt}}$ essentially determines the locations of these transitions.

For the optimization performed at $\eta_{\text{opt}}=0.26$, we see a very different progression upon reducing $\eta$, shown in Fig.~\ref{fgr:k30}. First, the pores at the optimized packing fraction are more anisotropic than in Fig.~\ref{fgr:k35}f with an average $\kappa^{2}$ value of 0.11 versus 0.05. The increased anisotropy is even more exaggerated at $\eta_{\text{opt}}=0.22$, where $\kappa^{2}=0.16$ at $\eta_{\text{opt}}$; see the Appendix. This is in addition to being more polydisperse (see Fig.~\ref{fgr:density}c), indicating overall reduced fidelity to the target simulation (which targeted monodisperse and spherical void spaces). Moreover, the structured progression of microphases observed for $\eta_{\text{opt}}=0.31$ above is absent; upon reducing $\eta$, the pores grow in size, accompanied by a modest increase in $\kappa^{2}$, until the void space percolates and $\kappa^{2}$ drops to zero. However, the growth in $\kappa^{2}$ corresponds to relatively unstructured and amorphous void spaces, not formation of columns, as can be seen in Fig.~\ref{fgr:k30}c,d. The analogous trend is observed for the particles, where percolated clusters span the space at high $\eta$ until they fragment into relatively anisotropic and polydisperse clusters. Taken together, it seems that the higher $\eta_{\text{opt}}$ conditions are more ideal for forming well-structured pores; however, pores formed at lower $\eta_{\text{opt}}$ have a larger stability range in $\eta$.  

The presence of distinct phase progressions upon varying $\eta$ is in agreement with prior work: increasing $\eta_{\text{opt}}$ enhances the magnitude of the features that constitute $u(r)$, analogous to lowering the temperature for a single pair potential with competing interactions. Previous studies have shown sharp structuring of mesophases and similar phase boundaries as a function of $\eta$ at low temperatures and more disordered regimes at higher temperatures.~\cite{SALR2,SALR3,lattice_bubbles_2} Moreover, the higher values of $\eta_{\text{opt}}$ that are needed to form well-structured pores are consistent with prior simulations in 3D where porosity was spontaneously observed for $\eta \approx 0.37$.~\cite{simulated_phases} 

\subsection{Porous Mesophases in Two Dimensions}

\begin{figure}[!htb]
  \includegraphics{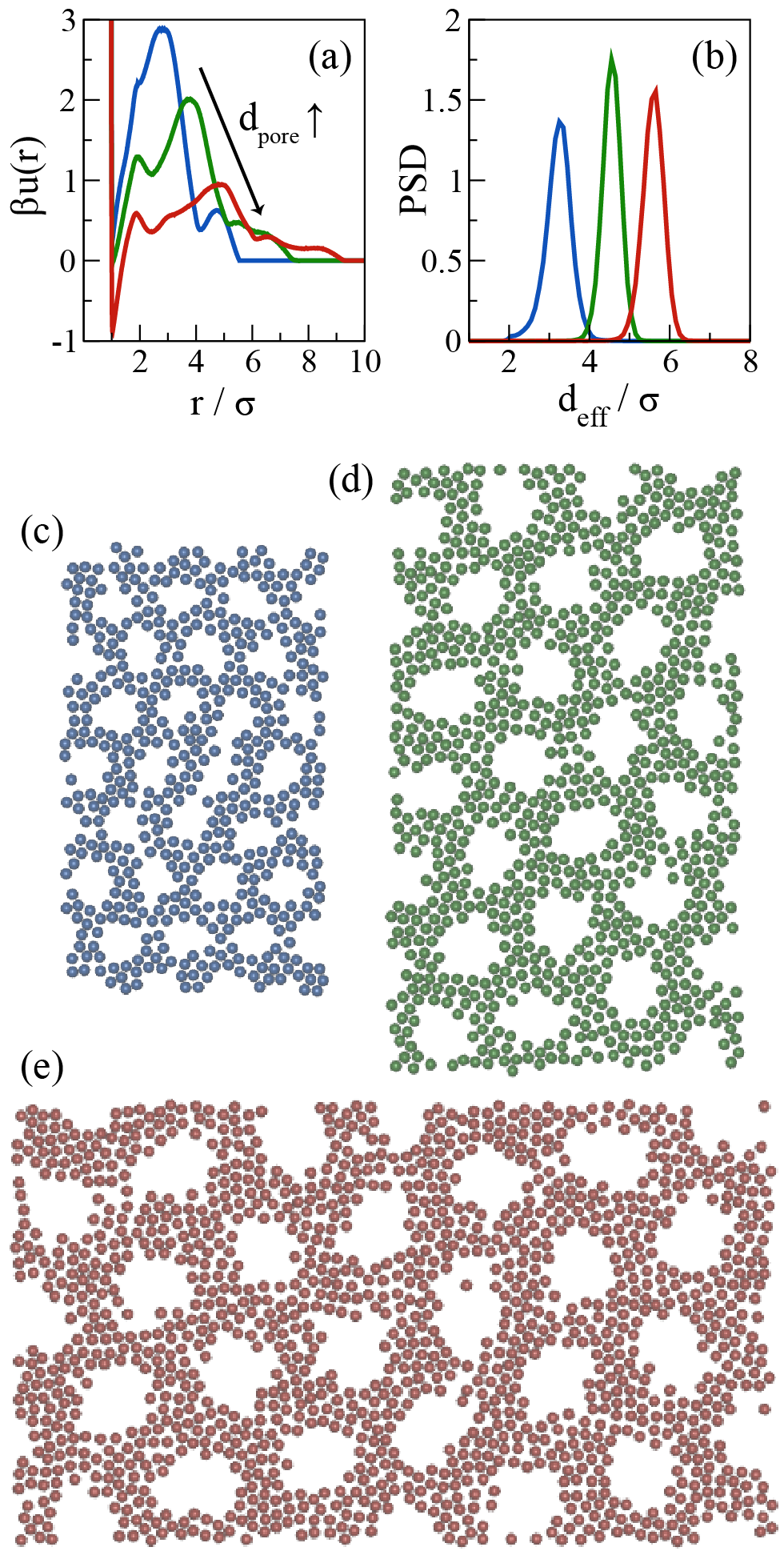}
  \caption{For $\eta_{\text{opt}}=0.60$, (a) dimensionless optimized pair potentials, $\beta u(r)$, (b) pore size distribution (PSD) functions, and (c-e) representative particle configurations for $d_{\text{pore}}=3$ (blue), 4 (green), and 5 (red).}
  \label{fgr:2D}   
\end{figure}

In addition to exploring the effect of dimensionality on pore-forming interaction potentials, the presence of pores in 2D, as opposed to 3D, might be easier to detect experimentally. 
In fact, a few 2D experimental systems possessing competing interactions have been reported to self-assemble into microphase-segregated objects.~\cite{jaime_expt,ghezzi_expt,AgNPs}
Therefore, we repeat the IBI procedure in 2D for $d_{\text{pore}}=$3, 4, and 5 with an empirically determined $\eta_{\text{opt}}$ of 0.60. The resultant potentials are shown in Fig.~\ref{fgr:2D}a; compared to 3D, the magnitudes of the features--particularly the repulsive elements--in the potentials are enhanced. Since fewer particles form the pore walls in 2D, each individual particle must contribute to a greater degree to build up a void space. The qualitative trend of the repulsive hump shifting to larger $r$ and decreasing in magnitude with increasing pore size mirrors the 3D result shown in Fig.~\ref{fgr:ps}a. The absolute depth of the attractive well grows with increasing $d_{\text{pore}}$; however, the difference between the maximum of the repulsive hump and the minimum of the attractive well actually decreases with pore size. Similarly to the main repulsive hump, the secondary, more subtle features of the potential are amplified relative to the 3D potentials--again, since the number of particles in each coordination shell is decreased in 2D, each particle must contribute to a greater degree to mimic the local environment of the target simulation.

\begin{figure}[!htb]
  \includegraphics{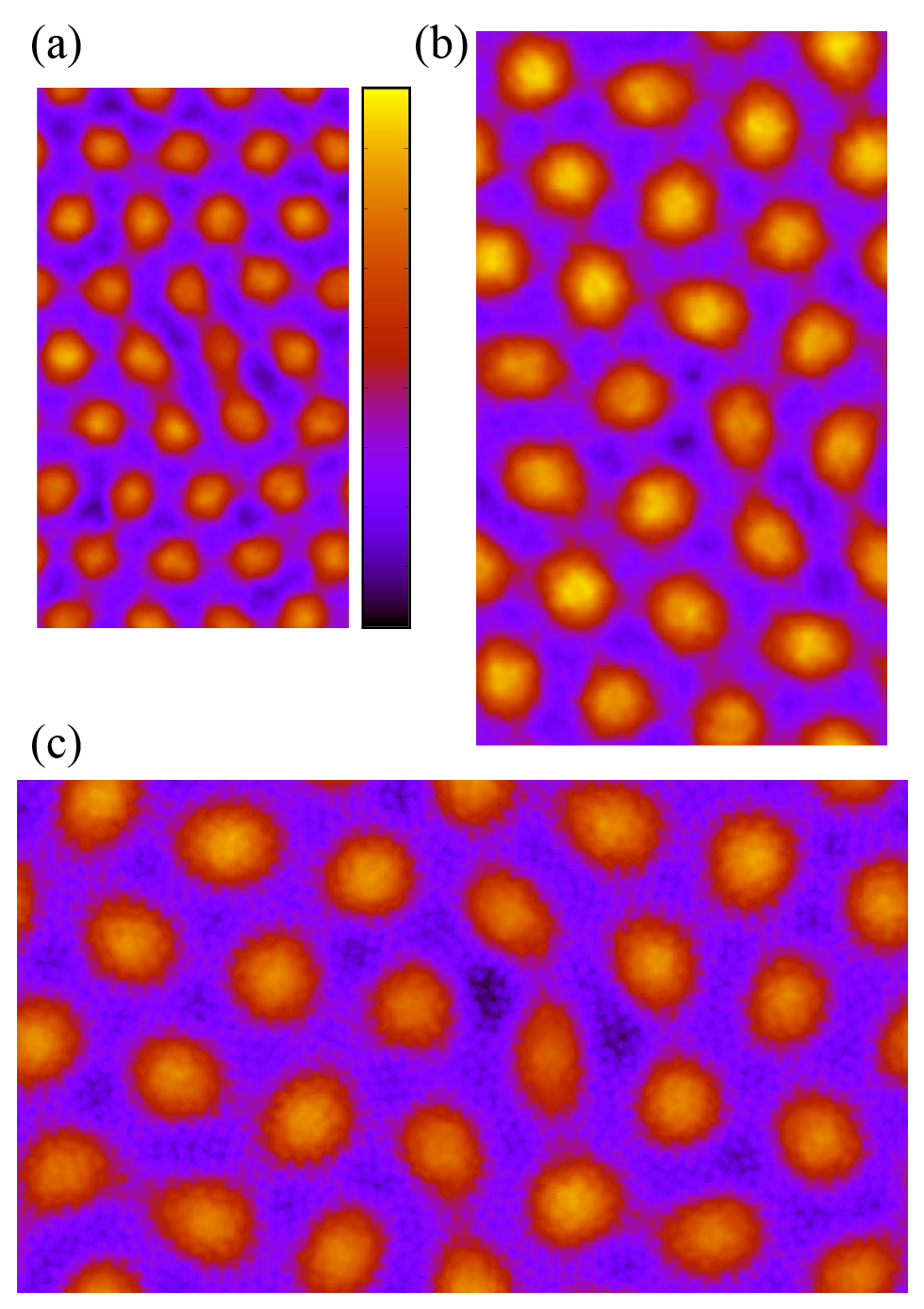}
  \caption{For $\eta_{\text{opt}}=0.60$ and $d_{\text{pore}}=3$ (a), 4 (b), and 5 (c), energy landscapes for insertion of a test particle, excluding the hardcore component of the pair potential. The ranges in energy (from black to yellow) are 35-70, 50-95, and 45-95 $k_{\text{B}}T$, respectively.}
  \label{fgr:energies}   
\end{figure}

As shown in the PSDs in Fig.~\ref{fgr:2D}b, the optimized potentials create reasonably monodisperse pores, though the resulting pores are somewhat larger than the targeted pores. Due to the differences in distance scalings in 2D compared to 3D, the ratio of nearest neighbor distance to pore size of 1.85 may be suboptimal for 2D pores. Representative configurations of the pore assemblies are shown in Fig.~\ref{fgr:2D}c-e. In all cases, the pores form the targeted triangular lattice. In Fig.~\ref{fgr:energies}, we show the energy landscape for the same configurations, omitting the hard-core contribution to the potential. For the void spaces, this quantity is the energy associated with insertion of a test particle; for the space occupied by particles, this metric reports on the environment in which the particles are embedded. In both the configurations and in the energy landscapes, it is clear that the pores for $d_{\text{pore}}=3$ are less well organized and separated than the larger pores. Due to the coupling of $d_{\text{pore}}$ and the width of the pore walls in the target simulation described in Sect.~\ref{subsec:pore_size}, a smaller prescribed pore size also thins the particle-rich regions. Though the particles move as a fluid in the simulation, some ordering is apparent in Fig.~\ref{fgr:2D}c due to the confinement imposed by the porous regions. At the same packing fraction, smaller pores necessitate more confined geometries and restricted configurational space for the occupied spaces.  Moreover, the average difference in energy between the occupied and unoccupied space is greater for $d_{\text{pore}}=3$ than for $d_{\text{pore}}=5$, 14.3 $k_{\text{B}}T$ versus 11.2 $k_{\text{B}}T$. While it is easy to imagine that increasing the void size would increase the entropic penalty for pore formation, these results indicate that the there is also an entropic penalty for smaller pores due to the imposed constraints on the interstitial particle positions. Ultimately, such considerations may limit the design of interactions that self-assemble into mesophases with pore diameters which approach $\sigma$.     

\section{Conclusions}
\label{sec:conclusions}

We have extended our prior work leveraging Iterative Boltzmann Inversion (IBI) to design porous mesophases with voids of a prescribed diameter to multiple pore sizes and packing fractions, allowing us to extract general trends as a function of these parameters (i.e., design rules). First, despite the different conditions explored, the potentials all have a general form that can be characterized by a relatively broad attractive well immediately following the hard-core-like component of the potential at $r=\sigma$. Beyond the attractive well in $r$ is a repulsive hump that peaks around the desired pore diameter and terminates around the pore nearest-neighbor distance. There are more subtle secondary features that appear to be related to the specifics of the target simulation and not essential to assembly of mesoscopic porous phases.  

By varying the pore diameter and the packing fraction of the target simulation, we find that the repulsive hump directly encodes, and is therefore highly sensitive to, the targeted pore size. Conversely, the packing fraction where the optimization is performed essentially dictates the depth of the attractive well (the associated breath is relatively constant, terminating at $\approx 2\sigma$ for all conditions). We also find that the potentials optimized at higher packing fractions ($\eta_{\text{opt}}=0.31$ and 0.35) possess greater structuring over a range in $\eta$: at $\eta_{\text{opt}}$, the pores are crystalline and, upon expansion of the simulation box, the pores evolve into other well-defined morphologies: inverse columns, lamellar sheets, columns of particles and finally clusters. By contrast, potentials optimized at lower packing fractions ($\eta_{\text{opt}}=0.22$ and 0.26) are mobile and more polydisperse at $\eta_{\text{opt}}$, and the pores become more amorphous without acquiring any notable higher order structuring upon expansion of the simulation box. 

In one sense, the pores associated with the potentials optimized at higher $\eta_{\text{opt}}$ are more ideal, in that they are more spherical and monodisperse. However, the pores also have a relatively small stability range in $\eta$, and they are also stationary within the simulation box due to their crystallinity. While the pores resulting from the lower $\eta_{\text{opt}}$ potentials are more polydisperse and do not organize into periodic structures, they display a more gradual transition as a function of $\eta$, with the pores slowly transforming into a percolated void space. Moreover, these pores are mobile, which may ultimately be a more desirable attribute, depending on the application.~\cite{pore_liquid_cages} Therefore, the optimal $\eta$ for forming pores will depend on which features of the pores (monodispersity, mobility, etc.) are most important for a given study. 

Finally, studying a range in pore diameter allows us to make some statements regarding the limits of this class of pore-forming potentials. The 2D calculations in particular demonstrated that, as the size of the pores approaches $\sigma$, the ability to form pores might be limited by entropic constraints; i.e, particle configurations are highly restricted by the presence of the voids. Application of our present methodology to larger sized pores is possible in principle, but presents practical challenges. For instance, to repeat these calculations for $d_{\text{pore}}=6$ at the packing fractions studied here would require a potential with a range of over 11$\sigma$ and a simulation box containing $\approx 15000-24000$ particles. Since hundreds of optimization cycles can be needed to achieve convergence and each cycle requiring a reasonably well-converged $g(r)$, these calculations would be computationally quite demanding.  

The computational expense of the IBI approach motivates future work using constrained potential forms that eliminate the interplay of the primary attractive well and repulsive hump with the more subtle, secondary features in the potential, potentially allowing for quantitatively and ultimately predictive, relationships between the characteristics of $u(r)$ and the resulting pores. Moreover, constraining the form of the pair interaction to physics-based models may provide insight into which types of experimental systems may be able to achieve these types of porous structures. 
As a starting point for such future work, ligand-coated nanoparticles appear to be an encouraging direction: interparticle interactions for Au nanoparticles coated with various charged alkanethiols have been computed that resemble the potentials reported in this work,~\cite{exp_pore_cluster_system} and Ag nanoparticles passivated with octanethiol have been observed experimentally to form clusters and stripe-like features at the air-water interface.~\cite{AgNPs} For the Au nanoparticles, decomposition of the interparticle interaction reveals that the hydrophobic effect associated with the ligands generates the attractive well, and electrostatic interactions provide a longer-ranged repulsive barrier. Therefore, physics-based models that represent these types of effects may be especially promising.~\cite{hydrophobic_effect,HansenMcDonald} Depletion forces are another obvious possibility for generating an attractive well,~\cite{HansenMcDonald} particularly to assemble smaller pores, and steric effects due to long ligands could potentially generate longer-ranged repulsions as well.~\cite{brush_potential} 


\renewcommand\thefigure{A\arabic{figure}}
\setcounter{figure}{0}
\renewcommand\thesection{A\arabic{section}.}
\setcounter{section}{0}

\section*{Acknowledgements}

This work was partially supported by the National Science Foundation (1247945) and the Welch Foundation (F-1696 and F-1848). S. D. gratefully acknowledges financial support from the S.N. Bose Scholars Program. We acknowledge the Texas Advanced Computing Center (TACC) at The University of Texas at Austin for providing HPC resources.

\section*{Appendix}

In this Appendix, we provide additional details about the Iterative Boltzmann Inversion calculations, pore size distributions (PSDs), anisotropy factors, and characterization of potentials for a pore diameter of 5$\sigma$.

\section{Comparison of Target and Optimized Radial Distribution Functions}

\begin{figure}[!htb]
  \includegraphics{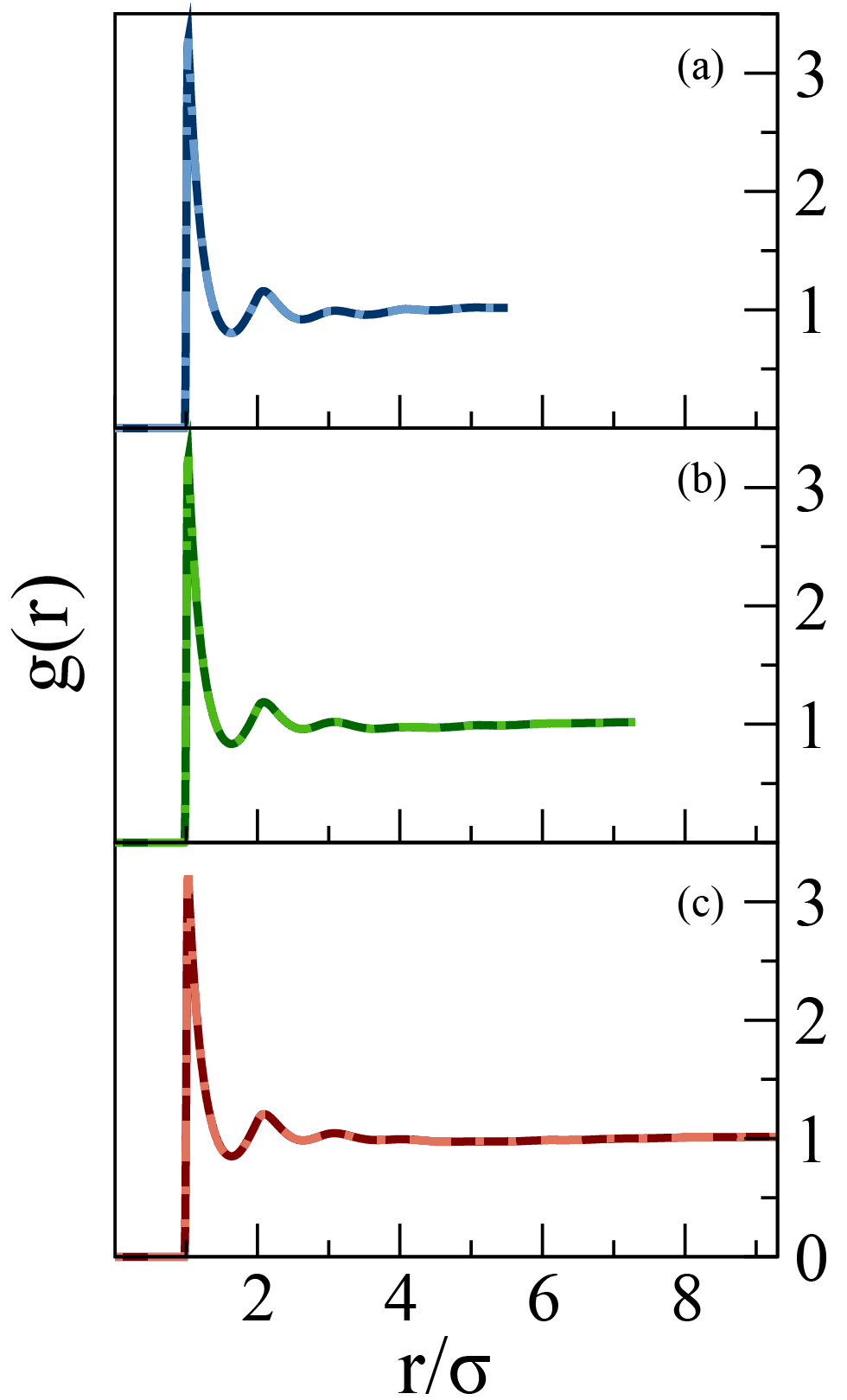}
  \caption{At $\eta_{\text{opt}}=0.31$, comparison of target (darker) and optimized (lighter) radial distribution functions for $d_{\text{pore}}=3$ (a), 4 (b), and 5 (c).}
  \label{fgr:rdfs}  
\end{figure}

In all cases, the matching between the radial distribution function of the target simulation and that of the optimized potential is excellent. Comparisons are shown in Fig.~\ref{fgr:rdfs} for the packing fraction of the small particles in the IBI optimization, $\eta_{\text{opt}}$, of 0.31 at all prescribed pore diameters ($d_{\text{pore}}$) examined in this work, but the results of all other optimizations are comparable in quality. As in our prior work, the effect of the pores on $g(r)$ is relatively subtle, primarily manifest as a slight dip in the $g(r)$ around the pore size.  

\section{Inserted Test Sphere Size Dependence}

\begin{figure}[!htb]
  \includegraphics{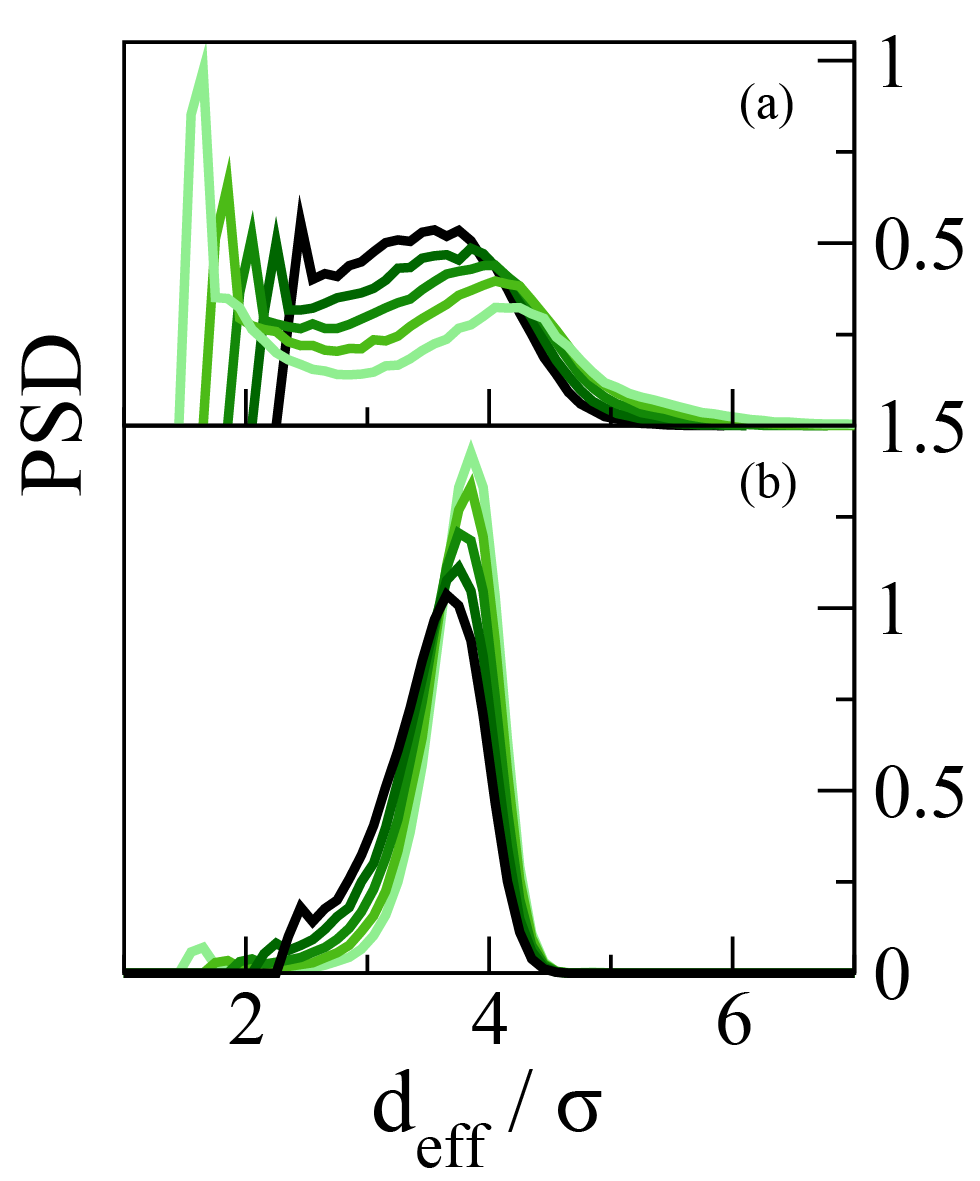}
  \caption{PSD functions for different inserted test sphere diameters (from light to dark: 1.6$\sigma$, 1.8$\sigma$, 2.0$\sigma$, 2.2$\sigma$, and 2.4$\sigma$) for $d_{\text{pore}}=4$ and $\eta_{\text{opt}}=$0.22 (a) and 0.31 (b).}
  \label{fgr:pstest}  
\end{figure}

As described in the main text, we analyze the pores by randomly inserting test spheres that do not overlap with the particles in the simulation. While the choice for the diameter of the test sphere is not unique, there are several considerations. For diameters that are too small, the test spheres will fill in both the pores and the naturally occurring interstitial spaces in the fluid. (This is somewhat analogous to a cluster size distribution for a clustered fluid, where both monomer and very small clusters are in coexistence with larger clusters of a preferred size.) As a result, the volume of the designed pores might be over estimated because the surrounding interstitial spaces might bleed into the the spherical pores. In the limit of infinite test points, all of the void space could be defined as a single pore, regardless of structure. On the other hand, if the test spheres are too large, then effects such as anisotropy and roughness of the pore surfaces might not be captured by the inserted test spheres, making the diameter of the pores artificially smaller. 

In Fig.~\ref{fgr:pstest}, we plot the PSDs for $d_{\text{pore}}=4$ and $\eta_{\text{opt}}=$ 0.22 (a) and 0.31 (b) with a variety of test sphere diameters. As expected, smaller inserted test spheres result in larger pores and vice versa. Throughout, we used test spheres with a diameter of 2$\sigma$ for the characterization; at approximately this diameter, the artificial peak in the PSD due to the discrete nature of the test spheres is minimized in both cases. As a complementary, though indirect, metric, we note that hard sphere-like simulations (using WCA particles, see Sect.~\ref{subsec:IBI} in the main text) performed at all $\eta_{\text{opt}}$ values indicated that spherical interstitial spaces of this size were extremely rare (data not shown). 

\begin{figure}[!htb]
  \includegraphics{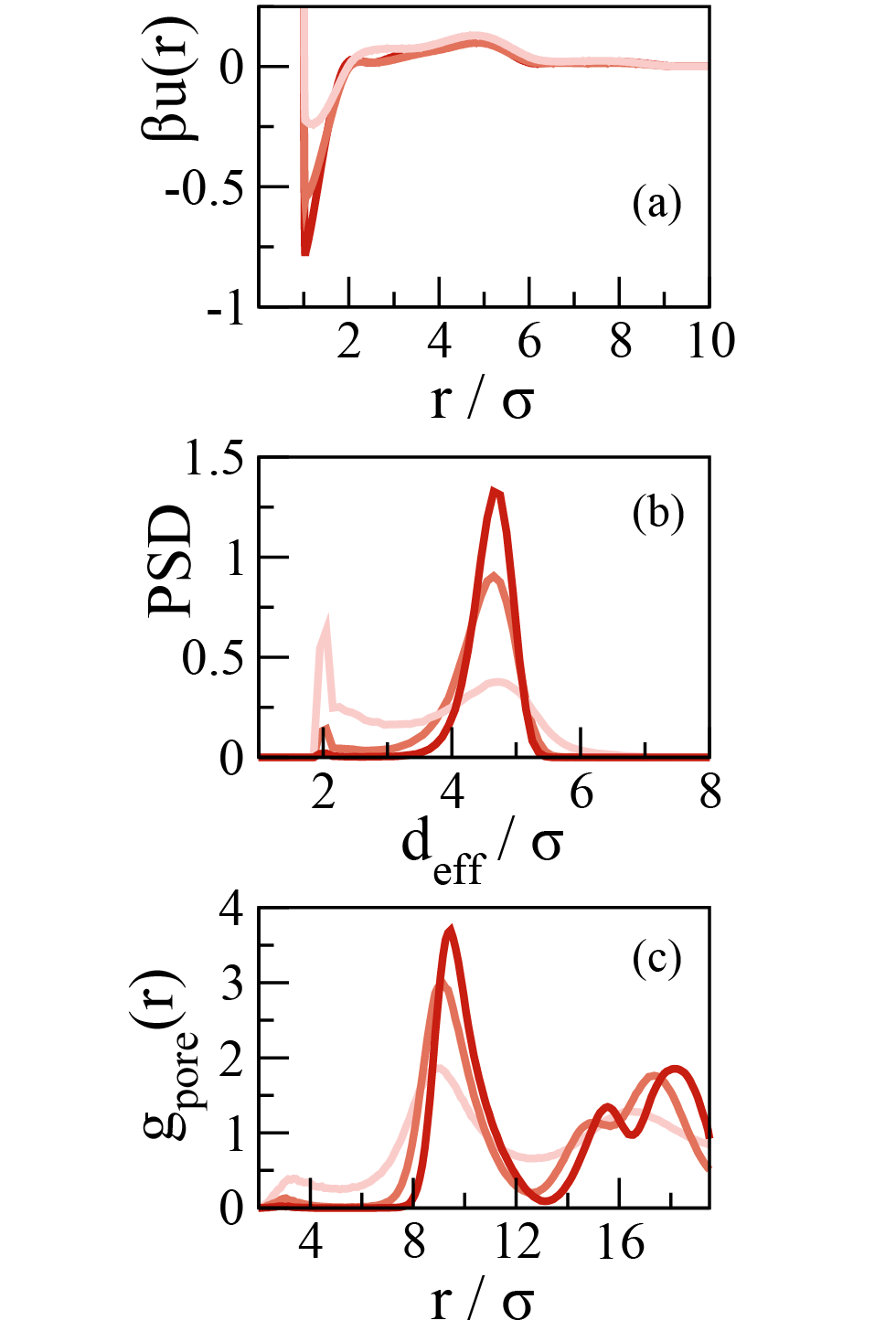}
  \caption{For $\eta$=0.22, 0.26, and 0.31 (from light to dark), (a) the dimensionless optimized pair potential, $\beta u(r)$, (b) PSD, and (c) $g(r)$ associated with the pore centers for $d_{\text{pore}}=5$.}
  \label{fgr:ps5}  
\end{figure}

\section{Analysis for $d_{\text{pore}}=5$ Target}

As shown in Fig.~\ref{fgr:ps5} for $\eta_{\text{opt}}=$0.22, 0.26, and 0.31, the results for $d_{\text{pore}}=5$ follow the same trends as $\eta_{\text{opt}}$ increases--increased amplitude of features in the potential, decreased polydispersity of the pores, and onset of pore crystallization--as those presented in the main text. 

\section{Additional Anisotropy Factor Calculations}

\begin{figure}[!htb]
  \includegraphics{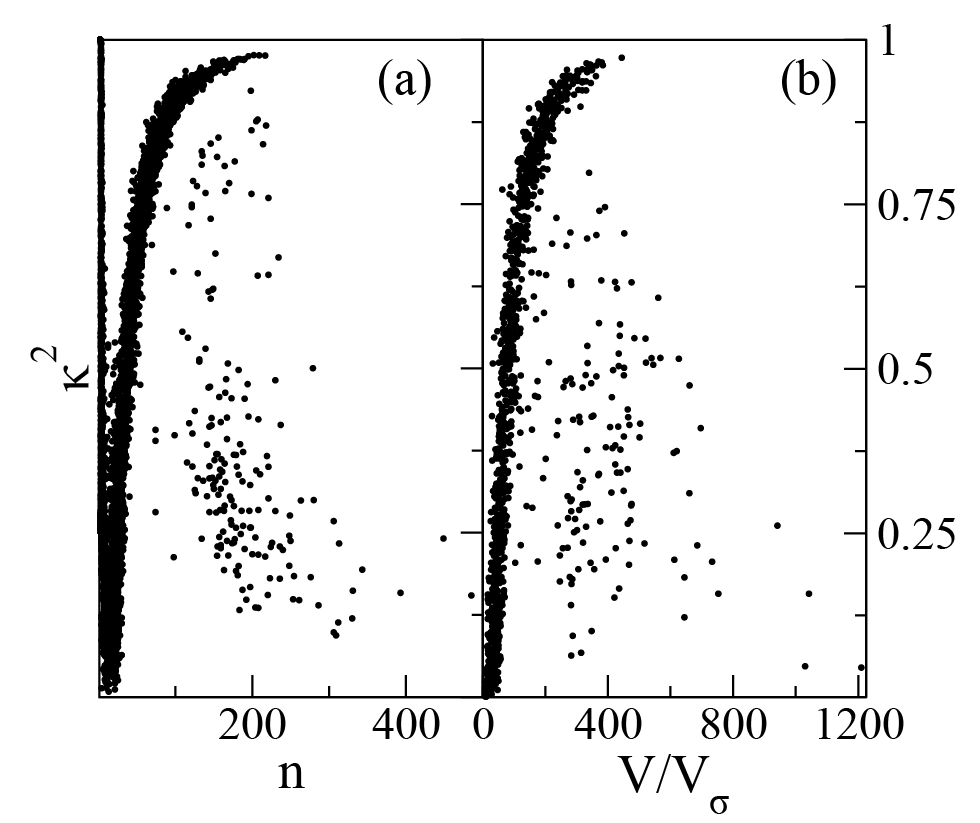}
  \caption{$\kappa^{2}$ as a function of (a) cluster size for the columnar phase shown in Fig. 3b in the main text and (b) pore volume (normalized by the volume of a particle with a diameter of $\sigma$) for the inverse columnar phase shown in Fig. 3d of the main text for 25 simulation snapshots.}
  \label{fgr:indk2}  
\end{figure}

\begin{figure}[!htb]
  \includegraphics{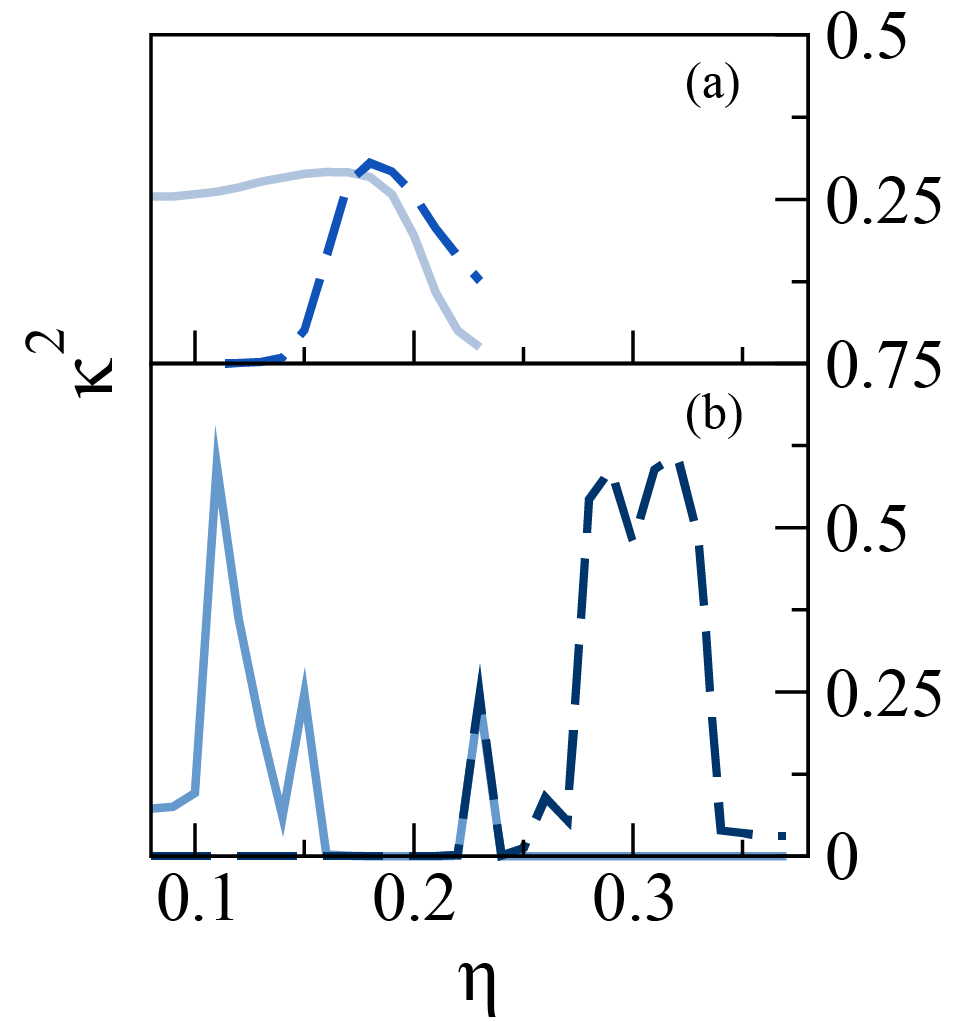}
  \caption{$\kappa^{2}$ as a function of $\eta$ for potentials optimized at packing fractions of (a) 0.22 and (b) 0.35.}
  \label{fgr:k2SI}  
\end{figure}

The $\kappa^{2}$ values shown in Fig.~\ref{fgr:k35}f and ~\ref{fgr:k30}f of the main text are averages for the discrete particle- or void-based objects, weighted by aggregate size and void volume, respectively. Fig.~\ref{fgr:indk2} shows $\kappa^{2}$ for the individual particle aggregates (left) and void regions (right) at packing fractions corresponding to particle column and void column phases, respectively. As anticipated in Sect.~\ref{subsec:char} of the main text, it is clear that small particle aggregates have highly variable anisotropy factors, and small voids are effectively spherical because they are typically comprised of only one or a few highly overlapping test sphere(s). Beyond this regime, the anisotropy is more or less directly related to the size of the feature--indicating that the columns (particle aggregate or void) are growing in length. However, for the larger features especially, there is a shift downward in $\kappa^{2}$, likely an artifact of multiple structures being identified as a single feature in the cluster analysis, a consequence of the large interfacial areas between columns. Such large features are heavily weighted in the average, which depresses the reported $\kappa^{2}$ values. However, from the data shown in Fig.~\ref{fgr:indk2}, it is clear that many columns do span the simulation box with the expected $\kappa^{2}$ values near to 1.

The analogous plots to Fig.~\ref{fgr:k35}f and ~\ref{fgr:k30}f for $\eta_{\text{opt}}=0.22$ and 0.35 are shown in Fig.~\ref{fgr:k2SI}. They follow straightforwardly from $\eta_{\text{opt}}=0.26$ and 0.31 shown in the main text. For $\eta_{\text{opt}}=0.22$, the particle-based curve is very similar to Fig.~\ref{fgr:k30}f, where the clusters are highly anisotropic (visual inspection reveals that they are amorphous as well). The void spaces are significantly more anisotropic than the $\eta_{\text{opt}}=0.26$ case, particularly at $\eta_{\text{opt}}$, indicating even greater discrepancies between the target simulation and the behavior of the optimized potentials as $\eta_{\text{opt}}$ is further reduced. On the other hand, as $\eta_{\text{opt}}$ is increased to 0.35, the microphases are again readily apparent, with columns of both particles and voids indicated by their relatively large $\kappa^{2}$ values at intermediate particle concentrations. 



\footnotesize{
\bibliographystyle{rsc} 
}

\providecommand*{\mcitethebibliography}{\thebibliography}
\csname @ifundefined\endcsname{endmcitethebibliography}
{\let\endmcitethebibliography\endthebibliography}{}

\end{document}